\documentstyle[aps,prb,twocolumn,epsf]{revtex}

\begin{document}
\title{VORTEX LATTICE MELTING AND THE DAMPING OF THE DHVA OSCILLATIONS IN THE MIXED
STATE }
\author{V.Zhuravlev$^{1}$,T.Maniv$^{1}$,I.D.Vagner$^{2}$, and P.Wyder$^{2}$}
\address{$^{1}$Chemistry Department, Technion-Israel Institute of\\
Technology, Haifa 32000, Israel\\
$^{2}$Grenoble High Magnetic Field Laboratory,Max-Planck-\\
Institute fur Festkorperforschung and CNRS, Grenoble Cedex 9,France.}
\date{\today}
\maketitle

\begin{abstract}
Phase fluctuations in the superconducting order parameter, which are
responsible for the melting of the Abrikosov vortex lattice below the mean
field $H_{c2}$ , are shown to dramatically enhance the scattering of
quasi-particles by the fluctuating pair potential , thus leading to enhanced
damping of the dHvA oscillations in the liquid mixed state. This effect is
shown to quantitatively account for the detailed field dependence of the
dHvA amplitude observed recently in the mixed state of a Quasi 2D organic SC.


%TCIMACRO{\TeXButton{vspace}{\vspace{.5in}  }}
%BeginExpansion
\vspace{.5in}  %
%EndExpansion
{}
\end{abstract}

Much effort has been recently invested in elucidating the fundamental
mechanisms in which the Abrikosov vortex lattice is involved in the damping
process of the de-Haas van-Alphen (dHvA) oscillations in the mixed state of
pure, type-II superconductors. Several theoretical attempts have been made
to account quantitatively for this effect (see discussions in Refs. \cite
{was98},\cite{zhur97} ). It seems , however, that non of the proposed
theories is capable of fully achieving this goal, in particular, since there
has been growing evidence recently to the significant sensitivity of the
dHvA amplitude to extrinsic factors such as vortex lattice disorder and flux
lines pinning\cite{tera97}, \cite{was98}. Furthermore, it has been shown
theoretically\cite{zhur97} that any mechanism destroying the phase coherence
in the vortex lattice, such as disorder or vortex lines fluctuations, should
dramatically enhance the inhomogeneous Landau level broadening of quasi
particles.

The effect of vortex lattice melting is of special interest since it is
associated with phase fluctuations of the superconducting order parameter,
which can be implemented selfconsistently into the Gorkov-Ginzburg-Landau
(GGL) formalism used in \cite{man92},\cite{zhur97}.

In a type-II superconductor under a magnetic field a soft shear Goldstone
mode, which can be described by long wavelength phase fluctuations \cite
{moo89} is connected to the Abrikosov lattice melting. \ Unfortunately
rigorous analytical approaches to this problem have encountered fundamental
difficulties: large order high temperature perturbation expansion with
Borel-Pade approximants to the low temperature behavior \cite{rug76}, \cite
{bre90} has no indication of an ordered vortex lattice even at zero
temperature. The existing non-perturbative approaches have not completely
clarified the situation: Renormalization group studies \cite{moo96} have
predicted no crystal vortex state in a pure 2D superconductor(SC) at finite
temperature, while the novel functional integral formalism suggested in Ref. 
\cite{tes94},\cite{her94}, has led to some kind of a vortex liquid freezing
transition without breaking the $U(1)$ symmetry. Several Monte Carlo
simulations, have recently shown \cite{kat93}$^{-}$\cite{sas94} that even in
a 2D SC a true vortex lattice melting phase transition takes place at finite
temperature and that the transition is of the first order.

In the present letter we make the first attempt to take into account the
effect of vortex lattice melting on the magnetization (dHvA) oscillations in
the mixed state. A simple, non-perturbative analytical model of melting is
developed for this purpose within the GGL theory for a pure 2D extremely
type II SC. Our model is based on the observation that in the vortex lattice
state the main correction to the mean field free energy arises from
fluctuating ''Bragg chains'', that is, from fluctuations which preserve long
range periodic order along some crystallographic direction in the vortex
lattice. This simplification reduces our 2D fluctuation problem to a 1D one,
which is then solved exactly.

Our starting point is the microscopic BCS Hamiltonian for electrons
interacting via an effective two-body attractive potential. We then write
down the formal functional integral expression for the partition function of
this system, eliminating the electronic field by introducing bosonic
Hubbard-Stratonovich complex field $\Delta \left( \overrightarrow{r}\right) $
, which describes all possible configurations of Cooper-pairs. By expanding
the resulting log determinant in the small $\Delta \left( \overrightarrow{r}%
\right) $, up to fourth order, we then recover a rather general expression
for the partition function in terms of the fully nonlocal Gorkov free energy
functional $F_{G}\left[ \left\{ \Delta \left( \overrightarrow{r}\right)
,\Delta ^{*}\left( \overrightarrow{r}\right) \right\} \right] $ \cite{tes94},%
\cite{man92} near $H_{c2}$:

\begin{equation}
Z=\int D\Delta \left( \overrightarrow{r}\right) D\Delta ^{\ast }\left( 
\overrightarrow{r}\right) \exp \left[ -F_{G}/k_{B}T\right]  \label{part}
\end{equation}

The stationary phase approximation for this functional integral yields the
mean field (GGL) equation for $\Delta \left( \overrightarrow{r}\right) $. In
the lowest Landau level approximation this equation can be solved by
expanding in a set of $\sqrt{N}$ Landau functions ($N$ being the total
number of vortices) $\phi _{n}\left( x,y\right) =\exp \left[
iq_{n}x-(y+q_{n}/2)^{2}\right] $ , with $q_{n}=\frac{2\pi }{a_{x}}n$ a wave
number along the $x$-axis, which is also proportional to the $y$-axis
position of a given $x$ -axis chain. Thus in the symmetric gauge 
\begin{equation}
\Delta \left( x,y\right) =e^{ixy}\sum\limits_{n=-\sqrt{N}/2}^{\sqrt{N}%
/2-1}a_{n}\phi _{n}\left( x,y\right)  \label{odpar}
\end{equation}
where for an arbitrary $2D$ periodic lattice the coefficients $a_{n}=\frac{%
\left( 2\pi \right) ^{1/4}}{a_{x}^{1/2}}\Delta _{0}\exp \left( i\gamma
n^{2}\right) $, \cite{man92}. All spatial variables are expressed in units
of the magnetic length. The parameter $\left| \Delta _{0}\right| ^{2}$
stands for the spatial average of the order parameter square, and $a_{x}$ is
the period along the $x$-axis. The summation in Eq.\ref{odpar} is performed
over $\sqrt{N}$ wavenumbers $q_{n}$ only. The reduction of the number of
degrees of freedom results from the assumed periodicity along the $x$ -axis.
Obviously $a_{x}$ is not unique and depends on the choice of the coordinate
system; any $2D$ lattice can be described by two different values of $a_{x}$%
, corresponding to the principal diagonals of the elementary cell (Fig.1).

In Refs.\cite{man92},\cite{zhur97} it was shown that the free energy
functional near stationary solutions can be approximated by a local
expression. This important property , resulting from gross cancellations of
nonlocal terms in the free energy due to destructive interference between
phase factors of the order parameter , simplifies the problem considerably ,
reducing it to the Ginsburg-Landau model with the well known local free
energy functional 
\begin{equation}
F_{GL}=\int d^{2}r\left[ -\alpha \left| \Delta \left( \overrightarrow{r}%
\right) \right| ^{2}+\frac{1}{2}\beta \left| \Delta \left( \overrightarrow{r}%
\right) \right| ^{4}\right]  \label{gl}
\end{equation}
Here $\alpha $ and $\beta $ are known function of temperature and magnetic
field (see below). It should be stressed, however, that the locality of the
free energy functional can be very sensitive to phase fluctuations,
especially in the vortex liquid state , since these fluctuations tend to
destroy the phase coherence of the order parameter. For the sake of
simplicity we ignore in this paper possible nonlocality in the Boltzman
factor of Eq.(\ref{part}). These assumption, though inconsistent with the
nonlocal form of the free energy in the vortex liquid state, can be shown by
a more careful analysis to have no significant influence on the final result.

It is convenient to express the GL free energy through the Abrikosov
parameter $\beta _{a}$: $F_{GL}=-N\varepsilon _{0}/\beta _{a}$, where $%
\varepsilon _{0}=\pi \alpha ^{2}/2\beta $. In the lattice state $\beta _{a}$
is a function of $a_{x}$ and $\gamma $. The dependence of $\beta _{a}$ on $%
z\equiv \pi ^{2}/a_{x}^{2}$ at $\gamma =\pi /2$ is plotted in Fig.(2a). Both
minima with $z_{1}=\pi /2\sqrt{3}$ and $z_{2}=\sqrt{3}\pi /2$ correspond to
the triangular lattice with $\beta _{a}\simeq 1,1596$. The maximum at $%
z_{3}=\pi /2$ is obtained for the square lattice\cite{kle63}.

In discussing the effect of fluctuations we first note that the order
parameter in Eq.\ref{odpar} with arbitrary coefficients $a_{n}$ includes
only those fluctuations which preserve periodic order along the $x$%
-direction. The number of independent coefficients $a_{n}$ is $\sqrt{N}$
whereas the total number of orbital centers is $N$. Therefore each $%
a_{n}=\left| a_{n}\right| e^{i\varphi _{n}}$ describes a set of $\sqrt{N}$
orbital centers periodically arranged within a certain chain along the $x$%
-axis (''vortex Bragg chain''). The phase $\varphi _{n}$ determines the
relative position $x_{n}=-$ $\varphi _{n}/q_{n}$ of the $n$-th chain.

The scale of the melting temperature is determined by the value of the phase
dependent terms in Eq.\ref{gl}. They can be readily calculated if we note
that in the GL Hamiltonian there is a small parameter $\lambda =\exp \left(
-z\right) $, with $z\gtrsim 1$ in the important regions near the minima of $%
\beta _{a}$ (see Fig.(2a)). Thus the quartic term can be written as an
expansion in $\lambda $, i.e. $\sum\limits_{n,s,t}\lambda
^{s^{2}+t^{2}}a_{n+s+t}^{\star }a_{n}^{\star }a_{n+s}a_{n+t}$, while the
leading phase dependent term is of the order $\lambda ^{2}$. Integrating
over amplitude fluctuation $\left| a_{n}\right| $ by using the stationary
phase approximation we find that the free energy can be written in the mean
field like form $\left\langle F_{GL}\right\rangle \simeq -N\varepsilon
_{0}/\beta _{a}$,where $\beta _{a}=\sqrt{\frac{z}{\pi }}\left[ 1+4\lambda
-4\lambda ^{2}\left\langle \cos \left( \chi _{n}\right) \right\rangle
_{phase}\right] $, $\chi _{n}$ $=\varphi _{n+1}+\varphi _{n-1}-2\varphi _{n}$
, and $\left\langle ...\right\rangle _{phase}$ means average over phase
fluctuations. Note that $\mu =\left\langle \cos \left( \chi _{n}\right)
\right\rangle _{phase}=\frac{I_{1}\left( T_{cm}/T\right) }{I_{0}\left(
T_{cm}/T\right) }$, where $\mu $ is the shear modulus of the Abrikosov
lattice normalized by its mean field value. Here $I_{k}\left( x\right) $ is
a modified Bessel function of the $k$-order, and $T_{cm}=\frac{4\lambda
^{2}\varepsilon _{0}}{\left( 1+4\lambda \right) \beta _{m}}$ is a
characteristic crossover temperature from the crystal to the liquid state,
with $\beta _{m}=\sqrt{\frac{z}{\pi }}\left( 1+4\lambda \right) $.Well below 
$T_{cm}$, $\mu \rightarrow 1$, so that thermal fluctuations do not distort
significantly the triangular lattice and the Abrikosov parameter is very
close to its minimal (mean field) value,$\beta _{A}$, i.e. $\beta _{a}\simeq
\beta _{A}\simeq \sqrt{\frac{z}{\pi }}\left( 1+4\lambda -4\lambda
^{2}\right) $. At $T\gg T_{cm}$ , on the other hand , $\mu \rightarrow 0$
and the vortex system transforms to a new (''liquid'') state for which $%
\beta _{a}\simeq \beta _{m}$. The crossover temperature $T_{cm}$ and the
corresponding free energy depend on the choice of vortex chains (Fig.1). At
zero temperature (i.e. in the vortex crystal state) the two minima at points 
$z_{1}$ and $z_{2}$ have the same energy (Fig.(2a)). At a finite temperature
this degeneracy is removed by the phase dependent terms, which introduce
different interchain coupling within different families of Bragg chains
(corresponding to $z_{1}$ and $z_{2}$). Since $z_{1}<z_{2}$, we have $%
T_{cm}\left( z_{1}\right) >T_{cm}\left( z_{2}\right) $. In Fig.(2b) we plot
local minimal values of the average free energy $\left\langle
F_{GL}\right\rangle $ as functions of the parameter $t=-\sqrt{\frac{%
4\varepsilon _{0}}{T}}$, which has been used in Ref.\cite{kat93}. It is
clear that at low temperatures the first state ( $z_{1}$ ) is more stable
than the second one ( $z_{2}$ ). Since $T_{cm}\left( z_{1}\right)
>T_{cm}\left( z_{2}\right) $ , the free energy of the first state increases
faster with increasing temperature than the second one, and so there is an
intersection point $T_{m}$ at which energies of both states are equal but
the corresponding entropies are a little different. Therefore we conclude
that at this point there is a weak first order transition characterized by a
small jump of the lattice entropy. The position of the crossing point $%
t_{m}\simeq -16$ and the jump of entropy $S=-T\frac{\partial F}{\partial T}$
at $T=T_{m}$: $T\Delta S/F_{MF}\simeq 7.5\times 10^{-3}$ , are in good
agreement with the Monte-Carlo simulations \cite{kat93}.

At higher temperatures $\beta _{a}$ does not describe correctly the free
energy. A straightforward calculation in this case yields for the partition
function per unit vortex: $Z_{v}\simeq \sqrt{\frac{2}{\beta }}e^{\beta
_{m}x^{2}}%
%TCIMACRO{\func{erf}}
%BeginExpansion
\mathop{\rm erf}%
%EndExpansion
\left( x\right) \exp \left[ -\lambda \left( \frac{\partial \ln 
%TCIMACRO{\func{erf}}
%BeginExpansion
\mathop{\rm erf}%
%EndExpansion
\left( x\right) }{\partial x}\right) ^{2}\right] $ ,where $x=sign\left(
\alpha \right) \frac{\sqrt{\varepsilon _{0}}}{\beta _{m}\sqrt{T}}$. In
contrast to mean field theory , in which there is a second order phase
transition,in our case there is only a crossover to the normal metal state,
with significant fluctuations of the superfluid density appearing at
magnetic fields far above the mean field $H_{c2}$.

All current theories of the dHvA effect in the mixed state are restricted to
the mean field approximation. The situation in the vortex liquid state is in
some sense similar to that described by Stephen\cite{mak91} in his
derivation of Maki's damping formula. However,in this approach a random
distribution of vortex lines was assumed, while the (nonzero) mean field
value of the Abrikosov lattice order parameter was used in the calculation.
The averaging over all realizations of the random vortex lattice in such a
model yields for the quartic term of the free energy: $\left\langle
a_{q+s+t}^{\star }a_{q}^{\star }a_{q+s}a_{q+t}\right\rangle \sim \left(
\delta _{s,0}+\delta _{t,0}\right) $. A similar expression, which reflects
the complete destruction of phase coherence in the vortex lattice, can be
naturally and consistently derived in the model discussed here at $T\gg T_{m}
$ by averaging over phase fluctuations, while at $T\gtrsim T_{m}$ it is
approximately valid.

Thus, similarly to the MS theory, our fluctuation theory predicts an
exponential damping: $M_{sc}=M_{n}\exp \left[ -\frac{\pi ^{3/2}}{n_{F}^{1/2}}%
\left\langle \left| \widetilde{\Delta }_{0}\right| ^{2}\right\rangle \right] 
$ , where $M_{sc}$, $M_{n}$ are amplitudes of the magnetization oscillations
in the superconducting and normal states respectively, $\widetilde{\Delta }%
_{0}=\Delta _{0}/\hbar \omega _{c}$, and $n_{F}=E_{F}/\hbar \omega _{c}$. A
simple analytic expression for the damping parameter can be derived in the
liquid state well above the melting point by integrating over amplitude
fluctuations ( after neglecting small linear corrections in $\lambda $): 
\begin{equation}
\left\langle \left| \Delta _{0}\right| ^{2}\right\rangle =\frac{1}{\pi }%
\frac{\partial \ln Z_{v}}{\partial \alpha }\simeq \frac{\alpha }{\beta }%
\left( 1+\frac{\partial \ln 
%TCIMACRO{\func{erf}}
%BeginExpansion
\mathop{\rm erf}%
%EndExpansion
\left( x\right) }{2x\partial x}\right)   \label{avop}
\end{equation}

This expression has no singularity at the mean field transition,where both $%
\alpha \rightarrow 0$ and $x\rightarrow 0$ , and smoothly interpolates
between the high field ($x\rightarrow -\infty $) value, $T/\pi \left| \alpha
\right| $ , and the low field ($x\rightarrow \infty $) mean field value, $%
\alpha /\beta $. We have compared this damping with the experimental results
on the organic quasi 2D SC: $\left( ET\right) _{2}Cu\left( SCN\right) _{2}$.
In this computation the coefficients $\alpha $ and $\beta $ derived
previously in Refs.\cite{man92},\cite{zhur97} for the GGL expansion have
been used, namely $\alpha =.5\pi \hbar \omega _{c}\ln \sqrt{H_{c2}/H}$, and $%
\beta =1.38\pi \hbar \omega _{c}/n_{F}$ , together with the well documented
Fermi surface and SC parameters of the studied material. The only adjustable
parameter in this scheme is the mean field $H_{c2}$. We have found that a
single value of about $\ 4.7\,\,T$ \ (see Fig.(3)) fits well the two
different sets of available experimental data, which were taken at quite
different temperatures \cite{wel95}, \cite{sas98}.

It should be stressed that in the fluctuation theory the magnetic
oscillations are smoothly damped well above $H_{c2}$ , and disappear below
the mean field $H_{c2}$ in remarkable quantitative agreement with
experiment. This contrasts the result of the mean field MS theory , where
the additional damping begins abruptly at $H_{c2}$ and then increases below $%
H_{c2}$ with significantly stronger rate than what is observed
experimentally,

The interesting point in our derivation of the four-particle correlation
function concerns the fact that its factorization in the high temperature
(or high field) limit into a product of two-particle correlation functions
is solely due to integration over phase fluctuations, which leads to the
limiting behavior:$\left\langle a_{q+s+t}^{\star }a_{q}^{\star
}a_{q+s}a_{q+t}\right\rangle _{phase}\sim $

$\left\langle \exp i\left( \varphi _{q+s}+\varphi _{q+t}-\varphi
_{q}-\varphi _{q+s+t}\right) \right\rangle _{phase}\rightarrow \left( \delta
_{s,0}+\delta _{t,0}\right) $ .

This is a generalized form of the condition for the vanishing of the shear
modulus $\mu $ in the liquid state. The apparent correlation between the
vanishing of the shear modulus of the Abrikosov lattice and the
factorization of the average phase factor of the quartic term in the SC free
energy is a general result, independent of the nature of the melting
transition. It is thus conceivable that at a first order melting transition,
the vanishing of $\mu $ at some finite value of the parameter $T/T_{m}$ , as
predicted by the Monte-Carlo simulation reported in\cite{sas94} ,will be
accompanied by a transition from 'coherent' \cite{zhur97} (i.e. relatively
weak) damping in the crystal state to 'incoherent' MS-like damping in the
liquid state at a certain magnetic field below $H_{c2}$.

Such a transition will be difficult to observe in the quasi $2D$ organic SC
investigated due to the very small value of the melting temperature $T_{m}$
and to the strength of the $2D$ fluctuations. In $3D$ superconductors where
the role of fluctuations is less important and the melting transition is
shifted significantly closer to $H_{c2}$ , one could expect such a
transition to be observable. It is thus interesting to note that in the
torque dHvA measurement performed on the borocarbide SC $YNi_{2}B_{2}C$\cite
{goll96} , where vortex lines pinning seemed to be weak (no peak effect near 
$H_{c2\text{ }}$) , the Dingle plot exhibited a rather sharp upward turn of
its (negative) slope at a certain field below $H_{c2}$ \cite{man98}, which
could indicate a freezing transition into an ordered vortex lattice state 
\cite{miock96}.

In conclusion, we have developed a simple analytical model of the vortex
lattice melting in a $2D$ type-II SC, which is in good quantitative
agreement with the state of the art numerical simulations in this field. We
have shown that fluctuations in the SC order parameter, which are
responsible for this melting, destroy the phase coherence in the quasi
particle scattering and thus lead to enhanced damping of the dHvA
oscillations in the mixed state, in remarkable quantitative agreement with
the experiment.

{\bf Acknowledgment:} We would like to thank Stephen Hayden and Mike
Springford for stimulating discussions of the experimental aspects of this
work. This research was supported by a grant from the US-Israel BSF, grant
no. 94-00243, and by the fund from the promotion of research at the Technion.

{\bf Figure Captions}

Fig.(1): Two different choices of the vortex chains in the triangular
lattice: dash lines - $z_{1}=\pi /2\sqrt{3}$, solid lines - $z_{2}=\sqrt{3}%
\pi /2$.

Fig.(2): (a) Mean field Abrikosov parameter,$\beta _{a}$, as a function of $%
z=(\pi /a_{x})^{2}$. (b) Dependence of the local minimal values of the free
energy $-\left\langle F\right\rangle /F_{MF}$,on the parameter $t=-2\sqrt{%
\varepsilon _{0}/T}$.

Fig.(3): The magnetization damping factor $R_{s}$ $=M_{sc}/M_{n}$, as a
function of magnetic field $H$ in the quasi $2D$ SC:$\left( ET\right)
_{2}Cu\left( SCN\right) _{2}$. Solid lines are our theoretical curves for $%
T=20mK,$and for the dHvA frequency $F=690T$ (1), and $T=120mK,F=600T$
(2);stars and circles are the corresponding experimental data taken from
Refs.\cite{wel95}, \cite{sas98}, respectively. The other parameters used in
the calculations are:$T_{c}=10.4K$ (from which $\Delta _{0}$ is calculated
using BCS weak coupling formula), and $m^{*}=3.5m_{e}$.

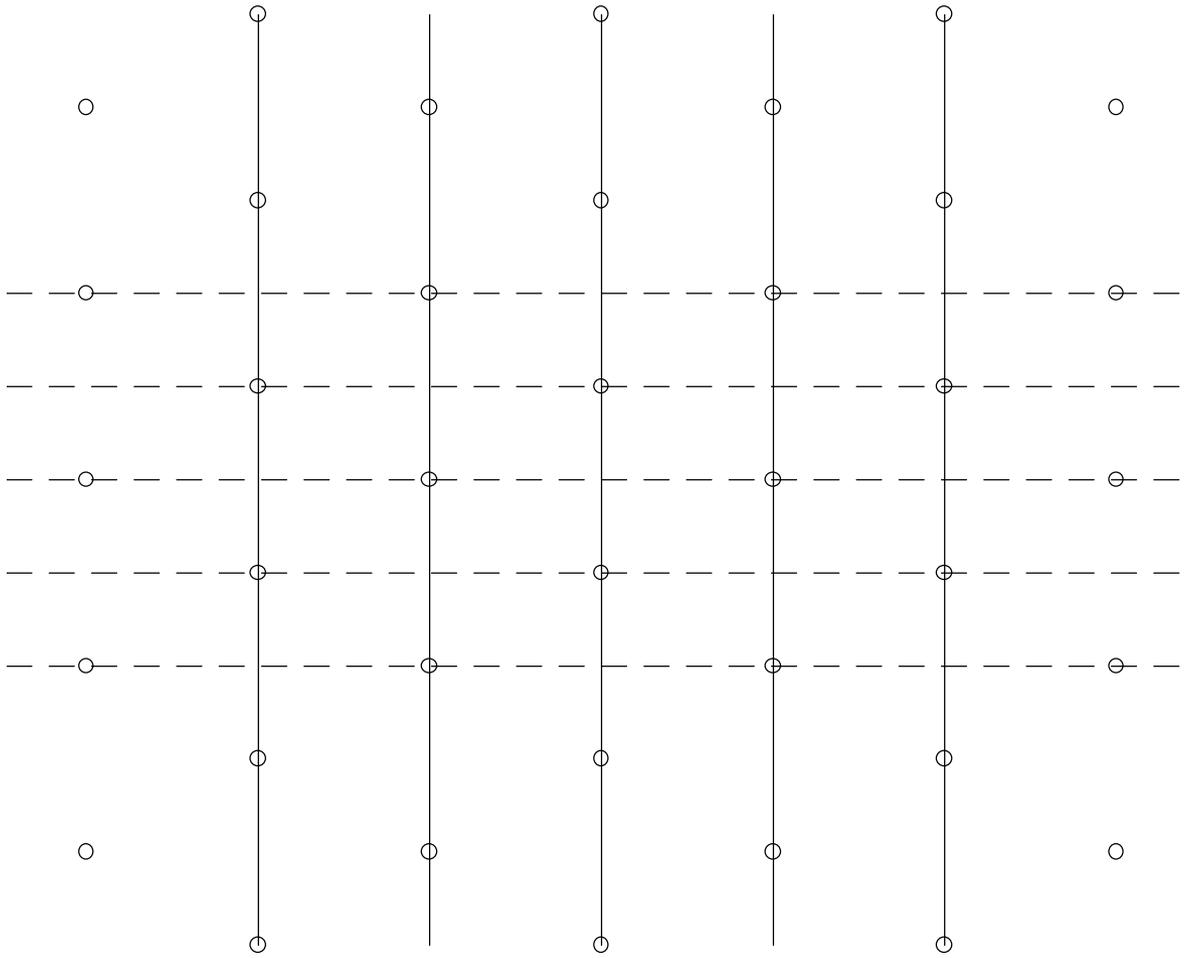
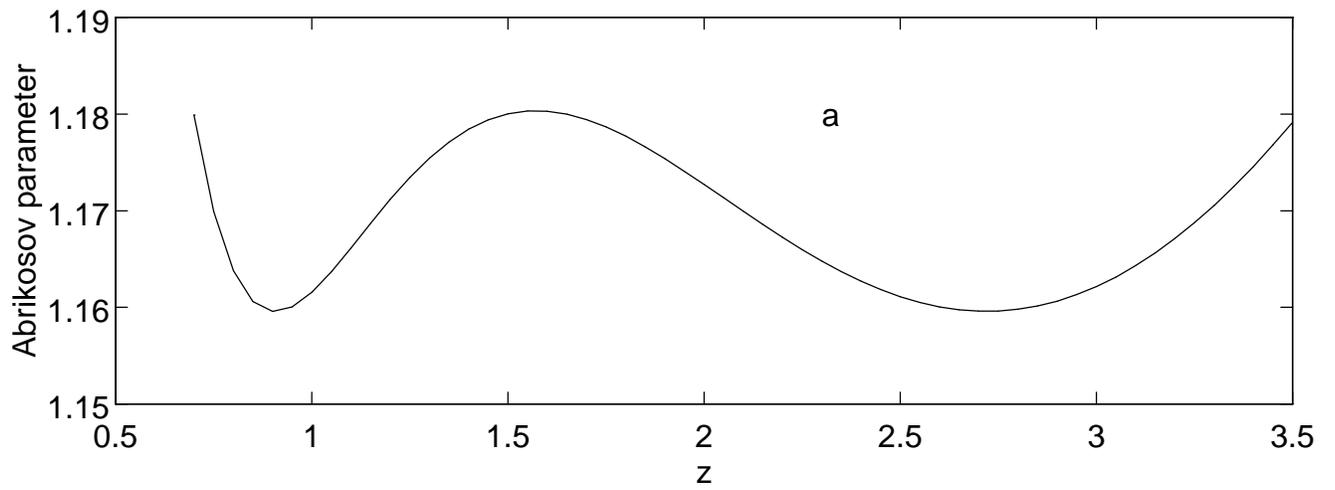
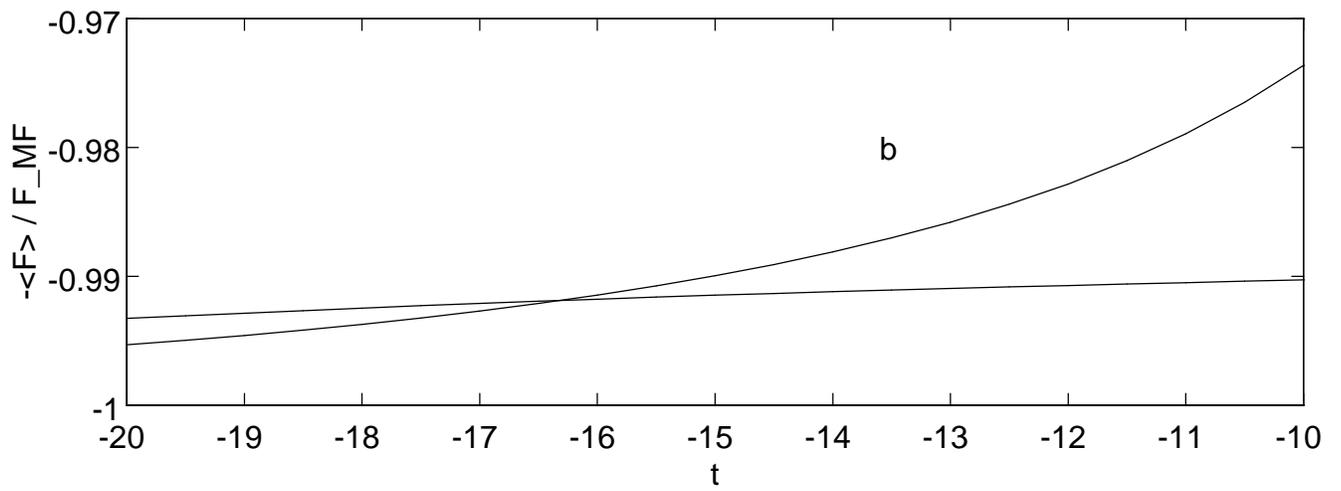
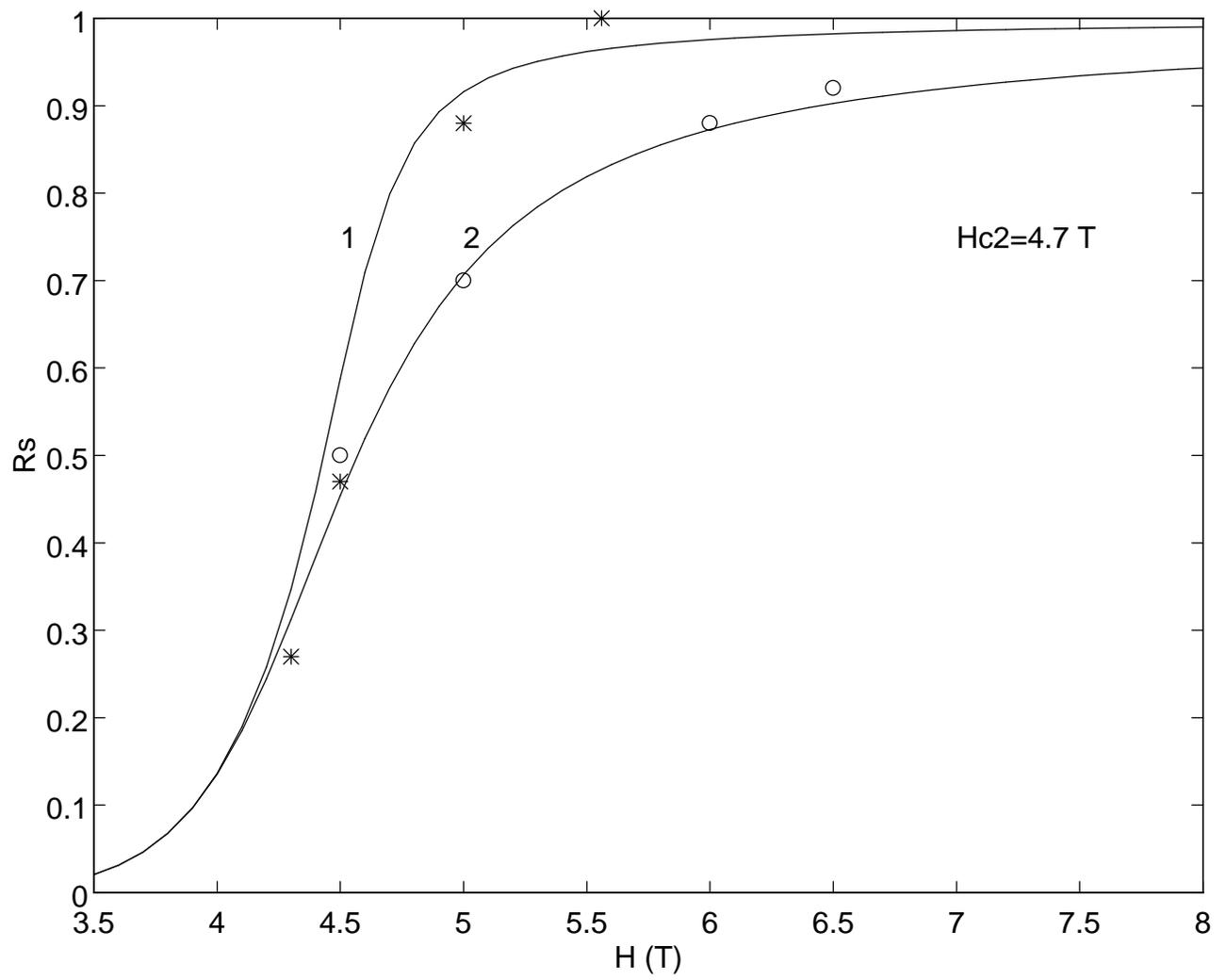
\end{document}